\documentclass{article}

\usepackage{fullpage}
\usepackage{charter}
\usepackage[american]{babel}
\usepackage{amsmath}
\usepackage{amssymb}
\usepackage{amsthm}
\usepackage{ulem}
\newtheorem{theorem}{Theorem}
\newtheorem{corollary}[theorem]{Corollary}
\newtheorem{definition}[theorem]{Definition}
\newtheorem{lem}[theorem]{Lemma}

\newcommand{\drop}[1]{}

\newcommand{\equals}{\stackrel{\mathrm{def}}{=}}

\title{Predictability of Fixed-Job Priority Schedulers \\on Heterogeneous Multiprocessor Real-Time Systems}
\author{Liliana Cucu-Grosjean
\and Jo\a"el Goossens}

\begin{document}
\maketitle

\begin{abstract}
  The multiprocessor Fixed-Job Priority (FJP) scheduling of real-time
  systems is studied. An important property for the schedulability
  analysis, the \emph{predictability} (regardless to the execution
  times), is studied for heterogeneous multiprocessor platforms. Our
  main contribution is to show that any FJP schedulers are predictable
  on \emph{unrelated} platforms. A convenient consequence is the fact
  that any FJP schedulers are predictable on \emph{uniform}
  multiprocessors.
\end{abstract}


\sloppy

\section{Introduction} \label{intro} 

A real-time system is often modelled as a finite collection of
independent recurring tasks, each of which generates a potentially
infinite sequence of \emph{jobs}. Every job is characterized by a
3-tuple $(r_{i},e_{i},d_{i})$, i.e., by a release time ($r_{i}$), an
execution requirement ($e_{i}$), and a deadline ($d_{i}$), and it is
required that a job completes execution between its arrival time and
its deadline.

From a theoretical and a practical point of view, we can distinguish (at
least) between three kinds of multiprocessor architectures (from less
general to more general):

\begin{description}
\item[Identical parallel machines] Platforms on which
all the processors are identical, in the sense that they have the same
computing power.

\item[Uniform parallel machines] By contrast, each processor in a
uniform parallel machine is characterized by its own computing
capacity, a job that is executed on processor $\pi_i$ of computing capacity
$s_i$ for $t$ time units completes $s_i \times t$ units of execution.

\item[Unrelated parallel machines] In unrelated parallel machines,
  there is an execution rate $s_{i,j}$ associated with each
  job-processor pair, a job $J_i$ that is executed on processor
  $\pi_j$ for $t$ time units completes $s_{i,j} \times t$ units of
  execution. 

\end{description}

In this paper, we consider real-time systems that are modeled by set of jobs and implemented upon a platform comprised of several \emph{unrelated} processors. We assume that the platform
\begin{itemize}
 \item is fully \emph{preemptive}: an executing job may be interrupted at any instant in time and have its execution resumed later with no cost or penalty.
 \item allows \emph{global} inter-processor migration: a job may begin
   execution on any processor and a preempted job may resume execution
   on the same processor as, or a different processor from, the one it
   had been executing on prior to preemption.
 \item forbids \emph{job parallelism}: each job is executing on at
   most one processor at each instant in time.
\end{itemize}

The \emph{scheduling algorithm} determines which job[s] should be
executed at each time-instant. Fixed-Job Priority (FJP) schedulers
assign priorities to jobs \emph{statically} and execute the
highest-priority jobs on the available processors. Dynamic Priority
(DP) schedulers assign priorities to jobs \emph{dynamically} (at each
instant of time). Popular FJP schedulers include: the Rate Monotonic
(RM), the Deadline Monotonic (DM) and the Earliest Deadline First
(EDF)~\cite{Liu}. Popular DP schedulers include: the Least Laxity
First (LLF) and the EDZL~\cite{thesisMok,Cho2005Efficient-Real-}.

The specified execution requirement of job is actually an upper bound
of its actual value, i.e., the worst case execution time (WCET) is
provided. The actual execution requirement may vary depending on the
input data, and on the system state (caches, memory, etc.). The
\emph{schedulability analysis}, determining whether all the jobs
\emph{always} meet their deadlines, is designed by considering a
finite number of (worst-case) scenarios (typically a single scenario)
assuming that the scheduler is \emph{predictable} with the following
definition: For a predictable scheduling algorithm, one may determine
an upper bound on the completion-times of jobs by analyzing the
situation under the assumption that each job executes for an amount
equal to the upper bound on its execution requirement; it is
guaranteed that the actual completion time of jobs will be no later
than this determined value.

\paragraph*{Related work.} Ha and Liu~\cite{Ha} `showed' that FJP
schedulers are predictable on \emph{identical} multiprocessor
platforms. However, while the result is correct, an argument used in the proof is not. Han and Park studied the predictability of the LLF
scheduler for identical multiprocessors~\cite{Han2006Predictability-}. To the best of our knowledge a single work addressed heterogeneous architectures, indeed we have showed in~\cite{etfaLCJG06} that any FJP schedulers are predictable on \emph{uniform} multiprocessors.

\paragraph*{This research.} In this research, we extend and correct~\cite{Ha} by considering unrelated multiprocessor platforms and by showing that any FJP schedulers are predictable on these platforms. 

\paragraph*{Organization.} The paper is organized as
follows. In Section~\ref{sec:model} we formally define the task
and machine models used, and provide some additional useful
definitions. In Section~\ref{sec:Ha} we show that the argument used by Han and Liu~\cite{Ha} is incorrect. In Section~\ref{sec:predic} we correct and extend the Ha and Liu result, we present our main contribution: the predictability of FJP schedulers on unrelated platforms. 

\section{Model and definitions}\label{sec:model}

We consider multiprocessor platforms $\pi$
composed of $m$ unrelated processors: $\{\pi_1,
\pi_2, \ldots, \pi_m \}$. Execution rates $s_{i,j}$ are
associated with each job-processor pair, a job $J_i$ that
is executed on processor $\pi_j$ for $t$ time units completes
$s_{i,j} \times t$ units of execution. For each job
$J_i$ we assume that the associated set of processors
$\pi_{n_{i,1}} > \pi_{n_{i,2}} > \cdots > \pi_{n_{i,m}} $
are ordered in a decreasing order of the execution rates
relative to the job: $s_{i,n_{i,1}} \geq s_{i,n_{i,2}}
\geq \cdots \geq s_{i,n_{i,m}}$. For identical execution
rates, the ties are broken arbitrarily, but consistently,
such that the set of processors associated with each job is
\emph{totally} ordered. For the processor-job pair $(\pi_j,J_i)$ if
$s_{i,j} \neq 0$ then $\pi_j$ is said to be an \emph{eligible} processor
for $J_i$.

\begin{definition}[Schedule $\sigma(t)$]\label{defSched}
  For any set of jobs $J \equals \{J_{1}, J_{2}, J_{3},\ldots\}$ and any
   set of $m$ processors $\{\pi_1, \ldots, \pi_m \}$ we define the
  \emph{schedule} $\sigma(t)$ of system $\tau $ at time-instant $t$ as
  $\sigma : \mathbb{N} \rightarrow \mathbb{N}^m$ where $\sigma(t) \equals (\sigma_1(t),
  \sigma_2(t), \ldots,
  \sigma_m(t))$ with \\
  $\sigma_j(t) \equals \left\{
\begin{array}{ll}
0, & \text{if there is no job scheduled on } \pi_j \text{at time-instant } t; \\
i, & \text{if job } J_i \mbox{ is scheduled on } \pi_j \text{ at time-instant } t.  
\end{array}
\right.$
\end{definition}

\begin{definition}[Work-conserving algorithm]\label{defWorkC} 
  An unrelated multiprocessor scheduling algorithm is said to be
\emph{work-conserving} if, at each instant, the algorithm schedules jobs
  on processors as follows: the highest priority (active) job $J_i$ is
  scheduled on its fastest (and eligible) processor $\pi_j$. The very
  same rule is then applied to the remaining active jobs on the
  remaining available processors.
\end{definition}

Throughout this paper, $J$ denotes a (finite or infinite) set of jobs: $J \equals \{
J_{1}, J_{2}, J_{3},\ldots\}$. We consider any FJP scheduler and without loss of generality we consider jobs in a decreasing order of
priorities $(J_1 > J_2 > J_{3} > \cdots$). We suppose that the
actual execution time of each job $J_i$ can be any value in the interval
$[e_i^{-}, e_i^{+}]$ and we denote by $J^{+}_i$ the job defined as
$J^{+}_i \equals (r_i,e_i^{+},d_i)$. The associated execution rates of
$J^{+}_i$ are $s_{i,j}^{+} \equals s_{i,j}, \forall j$.  Similarly,
$J^{-}_i$ is the job defined from $J_i$ as follows:
$J^{-}_i=(r_i,e_i^{-},d_i)$. Similarly, the associated execution rates
of $J^{-}_i$ are $s_{i,j}^{-} \equals s_{i,j}, \forall j$. We denote
by $J^{(i)}$ the set of the $i$ highest priority jobs (and its schedule by $\sigma^{(i)}$). We denote
also by $J^{(i)}_{-}$ the set $\{ J^{-}_1, \ldots, J^{-}_i \}$ and by
$J^{(i)}_{+}$ the set $\{J^{+}_1, \ldots, J^{+}_i \}$ (and its schedule by $\sigma^{(i)}_{+}$). Note that the
schedule of an ordered set of jobs using a work-conserving and
FJP scheduler is unique. Let $S(J)$ be the
time-instant at which the lowest priority job of $J$ begins its
execution in the schedule. Similarly, let $F(J)$ be the time-instant
at which the lowest priority job of $J$ completes its execution in the
schedule.

\begin{definition}[Predictable algorithm]\label{predAlg}
  A scheduling algorithm is said to be {\em predictable} if $S(J^{(i)}_{-})
  \leq S(J^{(i)}) \leq S(J^{(i)}_{+})$ and $F(J^{(i)}_{-}) \leq
  F(J^{(i)}) \leq F(J^{(i)}_{+})$, for all $1 \leq i \leq \#J$ and for all
schedulable $J^{(i)}_{+}$ sets of jobs.
\end{definition}

\begin{definition}[Availability of the processors]\label{defAvaiJob}
For any ordered set of jobs $J$ and any set of $m$ unrelated processors
$\{\pi_1, \ldots, \pi_m \}$, we define the
  {\em availability of the processors} $A(J,t)$ of the set of jobs $J$
  at time-instant $t$ as the set of available processors: $A(J,t) \equals
  \{j \mid \mbox{ } \sigma_j(t)=0 \} \subseteq \{1, \ldots, m \}$, where
  $\sigma$ is the schedule of $J$.
\end{definition}

\section{Proof from Ha and Liu~\cite{Ha}}\label{sec:Ha}

The result we extend in this work is the following:

\begin{theorem}
For any FJP scheduler and identical multiprocessor platform, the start time of every job is predictable, that is, $S(J^{(i)}_{-})
  \leq S(J^{(i)}) \leq S(J^{(i)}_{+})$.
\end{theorem}

We give here the first part of the original (adapted with our notations) proof of Ha and Liu, Theorem~3.1, page 165 of~\cite{Ha}.

\paragraph{Proof from~\cite{Ha}.} Clearly, $S(J^{(1)}) \leq S(J^{(1)}_{+})$ is true for the highest-priority job $J_{1}$. Assuming that $S(J^{(k)}) \leq S(J^{(k)}_{+})$ for $k<i$, we now prove $S(J^{(i)}) \leq S(J^{(i)}_{+})$ by contradiction. Suppose $S(J^{(i)}) > S(J^{(i)}_{+})$. Because we consider a FJP scheduler, every job whose release time is at or earlier than $S(J^{(i)}_{+})$ and whose priority is higher than $J_{i}$ has started by $S(J^{(i)}_{+})$ according to the maximal schedule $\sigma_{+}^{(i)}$. From the induction hypothesis, we can conclude that every such job has started by $S(J^{(i)}_{+})$ in the actual schedule $\sigma^{(i)}$. Because $e_{k} \leq e_{k}^{+}$ for all $k$, in $(0,S(J^{(i)}_{+}))$, the total demand of all jobs with priorities higher than $J_{i}$ in the maximal schedule $\sigma_{+}^{(i)}$ is larger than the total time demand of these jobs in the actual schedule $\sigma^{(i)}$. In $\sigma_{+}^{(i)}$, \uwave{a processor is available at $S(J^{(i)}_{+})$ for $J_{i}$ to start; a processor must also be available in $\sigma^{(i)}$ at or before $S(J^{(i)}_{+})$ on which $J_{i}$ or a lower-priority job can be scheduled}\ldots

\paragraph{Counter-example.} The use of the notion of \emph{``total
  demand''} used in the original proof is not appropriate considering
multiprocessor platforms. Consider, for instance, two set of jobs: $J
\equals \{J_{1} = J_{2} = (0,3,\infty)\}$ and $J' \equals \{J_{3} =
(1,5,\infty), J_{4} = (1,1,\infty)\}$. In $(0,2)$ the total demands of
both job sets are identical (i.e., 6 time units), if we schedule
the system using FJP schedulers, e.g., $J_{1}>J_{2}$ and $J_{3}>J_{4}$
it is not difficult to see that in the schedule of $J'$ a processor is
available at time 2 while in the schedule of $J$ we have to wait till
time-instant 3 to have available processor(s).
 
\section{Predictability}\label{sec:predic}

In this section we prove our main property, the predictability of FJP schedulers on unrelated multiprocessors which is based on the following lemma.

\begin{lem}\label{lemmaSoon} 
  For any schedulable ordered set of jobs $J$ (using a FJP and work-conserving scheduler) on an   arbitrary set of unrelated processors $\{\pi_1, \ldots, \pi_m\}$, we have $A(J^{(i)}_{+},t) \subseteq A(J^{(i)},t)$, for all $t$ and
  all $i$. In other words, at any time-instant the processors available in
  ${\sigma^{(i)}_{+}}$ are also available in ${\sigma^{(i)}}$. (We
  consider that the sets of jobs are ordered in the same decreasing
  order of the priorities, i.e., $J_1 > J_2 > \cdots > J_{\ell}$ and
  $J_1^{+} > J_2^{+} > \cdots > J_{\ell}^{+}$.)
\end{lem}
\begin{proof}
  The proof is made by induction by $\ell$ (the number of jobs).  Our
  inductive hypothesis is the following: $A(J^{(k)}_{+},t) \subseteq
  A(J^{(k)},t)$, for all $t$ and $1 \leq k \leq i$.
 
  The property is true in the base case since $A(J^{(1)}_{+},t)
  \subseteq A(J^{(1)},t)$, for all $t$. Indeed, $S(J^{(1)}) =
  S(J^{(1)}_{+})$. Moreover $J_{1}$ and $J_{1}^{+}$ are both scheduled on
  their fastest (same) processor $\pi_{n_{1,1}}$, but $J_{1}^{+}$ will
  be executed for the same or a greater amount of time than $J_{1}$.

  We will show now that $A(J^{(i+1)}_{+},t) \subseteq
  A(J^{(i+1)},t)$, for all $t$.

  Since the jobs in $J^{(i)}$ have higher priority than $J_{i+1}$, then
  the scheduling of $J_{i+1}$ will not interfere with higher priority jobs
  which have already been scheduled. Similarly, $J^{+}_{i+1}$ will not
  interfere with higher priority jobs of $J^{(i)}_{+}$ which have
  already been scheduled. Therefore, we may build the schedule
  $\sigma^{(i+1)}$ from $\sigma^{(i)}$, such that the jobs $J_1, J_2,
  \ldots, J_{i}$, are scheduled at the very same instants and on the
  very same processors as they were in $\sigma^{(i)}$. Similarly, we
  may build $\sigma^{(i+1)}_{+}$ from $\sigma^{(i)}_{+}$.

  Note that $A(J^{(i+1)},t)$ will contain the same available
  processors as $A(J^{(i)},t)$ for all $t$ except the time-instants at
  which $J^{(i+1)}$ is scheduled, and similarly $A(J^{(i+1)}_{+},t)$
  will contain the same available processors as $A(J^{(i)}_{+},t)$ for
  all $t$ except the time-instants at which $J^{(i+1)}_{+}$ is
  scheduled. From the inductive hypothesis we have $A(J^{(i)}_{+},t)
  \subseteq A(J^{(i)},t)$, we will consider time-instant $t$, from
  $r_{i+1}$ to the completion of $J_{i+1}$ (which is actually not
  after the completion of $J_{i+1}^{+}$, see below for a proof), we
  distinguish between four cases:
\begin{enumerate}
\item $ A(J^{(i)},t) = A(J^{(i)}_{+},t)=0$: in both situations no processor
  is available. Therefore, both jobs, $J_{i+1}$ and $J_{i+1}^{+}$, do
  not progress and we obtain $A(J^{(i+1)},t) = A(J^{(i+1)}_{+},t)$. The
  progression of $J_{i+1}$ is identical to $J_{i+1}^{+}$.

\item \label{item:add} $A(J^{(i)},t) \neq \emptyset$ and
  $A(J^{(i)}_{+},t) = \emptyset$: if an eligible processor exists, $J_{i+1}$ progress on an available processor in $A(J^{(i)},t)$ not available in $A(J^{(i)}_{+},t)$, $J_{i+1}^{+}$ does not progress. Consequently, $A(J^{(i+1)}_{+},t) \subseteq A(J^{(i+1)},t)$ and the progression of $J_{i+1}$ is
  strictly larger than $J_{i+1}^{+}$.
\item $\label{item:idem} A(J^{(i)},t) = A(J^{(i)}_{+},t) \neq
  \emptyset$: if an eligible processor exists, $J_{i+1}$ and
  $J_{i+1}^{+}$ progress on the same processor. Consequently,
  $A(J^{(i+1)}_{+},t) = A(J^{(i+1)},t)$ and the progression of
  $J_{i+1}$ is identical to $J_{i+1}^{+}$.
\item $A(J^{(i)},t) \neq A(J^{(i)}_{+},t) \neq \emptyset$: if the faster processor in $A(J^{(i)},t)$ and $A(J^{(i)}_{+},t)$ is the same see previous case of the proof (case~\ref{item:idem}); otherwise 
$J_{i+1}$ progress on a \emph{faster} processor than $J_{i+1}^{+}$, that processor is not available in $A(J^{(i)}_{+},t)$, consequently a slower processor remains idle in $A(J^{(i)},t)$ but busy in $A(J^{(i)}_{+},t)$. Consequently, $A(J^{(i+1)}_{+},t) \subseteq A(J^{(i+1)},t)$ and the progression of $J_{i+1}$ is larger than $J_{i+1}^{+}$.
\end{enumerate}
Therefore, we showed that $A(J^{(i+1)}_{+},t) \subseteq
A(J^{(i+1)},t)$ $\forall t$, from $r_{i+1}$ to the completion of
$J_{i+1}$ and that $J_{i+1}$ does not complete after
$J_{i+1}^{+}$. For the time-instant after the completion of $J_{i+1}$ the
property is trivially true by induction hypothesis.
\end{proof}

\begin{theorem} \label{thNotWorkPred} FJP schedulers are predictable on unrelated platforms.
\end{theorem}
\begin{proof}
In the framework of the proof of Lemma~\ref{lemmaSoon} we actually showed extra properties which imply that FJP schedulers are predictable on unrelated platforms: (i) $J_{i+1}$ completes not after $J_{i+1}^{+}$ and (ii) $J_{i+1}$ can be scheduled either at the very same instants as $J_{i+1}^{+}$ or may progress during \emph{additional} time-instants (case~(\ref{item:add}) of the proof) these instants may precede the time-instant where $J_{i+1}^{+}$ commences its execution.
\end{proof}

Since multiprocessor real-time scheduling theory is a relative new area of research, researchers pay much more attention to uniform platforms (at least nowadays), we find convenient to emphasized the following corollary:

\begin{corollary}
FJP schedulers are predictable on uniform platforms.
\end{corollary}

\section{Conclusion}

We extended and corrected~\cite{Ha} by considering unrelated multiprocessor platforms and by showing that any FJP schedulers are predictable. A convenient consequence is the fact that any FJP schedulers are predictable on \emph{uniform} multiprocessors. 

These results constitute a first theoretical foundation in order to design efficient schedulability tests for the scheduling of periodic and sporadic real-time tasks on heterogeneous architectures.

\bibliographystyle{acm}
\bibliography{biblio.bib}

\end{document}